\documentclass[aps,prl,twocolumn]{revtex4-1}
\usepackage{hyperref}
\hypersetup{colorlinks=true, citecolor=blue, urlcolor=blue, linkcolor=blue}

\pdfoutput=1 
\usepackage[utf8]{inputenc}
\usepackage[T1]{fontenc}
\usepackage{amssymb}
\usepackage{amsmath}
\usepackage{graphics}
\usepackage{color}
\usepackage{calc}

\renewcommand{\d}{\textrm{d}}
\renewcommand{\vec}[1]{\mathbf{#1}}
\newcommand{\e}{{\mathrm{e}}}
\newcommand{\im}{\mathrm{i}}
\newcommand{\diagramBox}[2][0.5]{\raisebox{0.5ex-#1\height}{#2}}

\newcommand{\Eq}[1]{Eq.~\eqref{#1}}
\newcommand{\Fig}[1]{Fig.~\ref{#1}}

\begin{document}

\title{
  On the duality of ring and ladder diagrams and its importance for
  many-electron perturbation theories
}

\author{Andreas Irmler}
\author{Felix Hummel}
\author{Andreas Gr\"uneis}
\email{andreas.grueneis@tuwien.ac.at}
\affiliation{
  Institute for Theoretical Physics, TU Wien,\\
  Wiedner Hauptstraße 8-10/136, 1040 Vienna, Austria
}
\date{\today, PREPRINT}


\begin{abstract}
We present a diagrammatic decomposition of the
transition pair correlation
function for 
the uniform electron gas.
We demonstrate explicitly that ring and ladder diagrams
are dual counterparts that capture significant long- and short-ranged
interelectronic correlation effects, respectively.
Our findings help to guide
the further development of approximate many-electron theories
and reveal that the contribution of the ladder diagrams to the electronic
correlation energy can be approximated in an effective manner using second-order
perturbation theory.
We employ the latter approximation to reduce the
computational cost of coupled cluster theory calculations
for insulators and semiconductors by two orders of
magnitude without compromising accuracy.

\end{abstract}

\maketitle


\emph{Introduction.} -- The ongoing advancement of modern electronic structure theories
over the last few decades 
led to a proliferation of 
computational studies and research on atoms,
molecules and condensed matter systems based on first principles.
Electronic structure methods allow to simulate
a wide range of
physically as well as chemically relevant material properties including
thermodynamic stability, electric conductivity,
magnetic ordering and optical properties.
Yet many of the most successful methods such
as approximate density functional theory or
many-electron perturbation theories rely substantially
on fortuitous error cancellation with far-reaching consequences.
Attempting to improve upon individual terms in the
many-body perturbation expansion of the electronic exchange and correlation
energy by employing higher levels of theory often has a contrary
effect and deteriorates the achieved level of accuracy,
making it hard to approach the exact
result in a systematic manner.
A profound physical understanding of the
individual perturbation theory contributions to the ground- and excited state
energies of many-electron systems and their balance is, however,
needed to design novel, more efficient and concomitantly
more accurate theories.
Here, we present a new method that makes it possible
to decompose the many-body contributions to the
correlation energies of \emph{ab-initio} systems
as well as the uniform electron gas in a manner that enables
a deeper understanding of widely-used approximations.
Furthermore we demonstrate that the obtained findings can be used to 
improve the efficiency of many-electron theory calculations.

In many-body physics and quantum chemistry,
approximate methods are often tailored to become exact
in limiting cases.
Examples for such electronic structure theories include:
(i) the random-phase approximation (RPA),
which captures the leading order contributions to the
correlation energy of the uniform electron gas
in the high density limit~\cite{macke_uber_1950,pines_collective_1952,gell-mann_correlation_1957},
(ii) the ladder theory which works reliable in the low-density limit~\cite{Freeman_1983,yasuhara_1972,yasuhara_1974},
and
(iii) coupled cluster singles and doubles theory (CCSD) that combines the
latter two approaches and is exact
for two-electron systems~\cite{coester_short-range_1960,cizek_use_1969}.
Yet all of the methods mentioned above exhibit
shortcomings for real materials, respectively:
systematic overcorrelation,
divergence of the correlation energy for metals,
and a poor trade-off
between computational cost and accuracy.
In a remarkable manner, however, errors like the overcorrelation
of the RPA, cancel out when considering energy differences,
and sometimes even mimic more sophisticated electronic
correlation effects~\cite{henderson2010,Harl2009,Dobson2006}.
Furthermore disregarding exchange-like
interactions between electron pairs in coupled cluster
theory systematically improves upon the achieved level
of accuracy~\cite{kats_2013}.
These examples illustrate that it is imperative to fully understand
the effect of individual contributions as a function of the electronic
density and interelectronic distance
in order to guide the further development of more accurate and
efficient approximations to the electronic correlation energy
in strongly as well as weakly correlated systems.

\emph{Theory.} -- In this work we introduce a decomposition of the electronic
transition pair correlation function and the corresponding correlation
energy contributions.
For the sake of brevity we will restrict the following discussion 
to the uniform electron
gas (UEG) and turn to \emph{ab-initio} systems afterwards.
The correlation energy can be written as an integral of the Coulomb
potential $v(\vec r_{12})=1/|\vec r_{12}|$ and a function
$g(\vec r_{12})$ of the inter-electronic radius $\vec r_{12}$
\begin{equation}
E_\mathrm{c} = \int\d\vec r_{12}\, g(\vec r_{12}) v(\vec r_{12}).
\label{eq:ecorr}
\end{equation}
The function $g(\vec r_{12})$ is called transition pair correlation function
(PCF). It can be explicitly given by
\begin{equation}
g(\vec{r}_{12}) = \frac1{(2\pi)^3}
  \sum_{abij}\e^{\im  \left (\vec k_i-\vec k_a
\right )\cdot  \vec r_{12}} (2t_{ij}^{ab}-t_{ji}^{ab})
\label{eq:gofr}
\end{equation}
and corresponds to the Fourier transform of the transition structure factor
$S(\vec q)$, recently studied in Refs.~\cite{Liao2016,Gruber2018}.
The indices $i$, $j$ and $a$, $b$ label occupied and virtual
spatial orbitals, respectively. In the UEG, orbitals correspond to
plane waves with respective wave vectors $\vec k_i$, $\vec k_j$ and
$\vec k_a$, $\vec k_b$.
The amplitudes $t_{ij}^{ab}$ are obtained by solving
the underlying amplitude equations on the respective level
of  many-electron theory such as second-order perturbation theory~\cite{moller_note_1934},
coupled cluster singles and doubles (CCSD) theory~\cite{coester_short-range_1960,cizek_use_1969},
the random-phase approximation~\cite{macke_uber_1950,pines_collective_1952,gell-mann_correlation_1957} or
ladder theory~\cite{yasuhara_1972,yasuhara_1974}.
In coupled cluster doubles theory, the $t_{ij}^{ab}$'s are obtained by
solving the recursive amplitude equations \cite{szabo_modern_1996,shavitt_2009} that include
all terms also present in the ring and the ladder approximation
\begin{widetext}
\begin{equation}
\begin{array}{r c rc cc cc cc cc cc cc ccc}
  &
  &
  & \overbrace{\rule{6ex}{0pt}}^\text{driver}
  &
  & \multicolumn{3}{c}{\overbrace{\rule{24ex}{0pt}}^\text{ring}}
  &
  & \overbrace{\rule{8ex}{0pt}}^\text{ppl}
  &
  & \multicolumn{5}{c}{\overbrace{\rule{34ex}{0pt}}^\text{rest}}
  \\
    \diagramBox[1]{\includegraphics{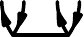}}
  & =
  &
  & \diagramBox[1]{\includegraphics{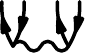}}
  & +
  & \diagramBox[1]{\includegraphics{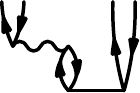}} 
  & +
  & \diagramBox[1]{\includegraphics{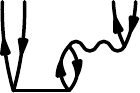}}
  & +
  & \diagramBox[1]{\includegraphics{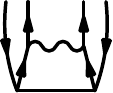}}
  & +
  & \diagramBox[1]{\includegraphics{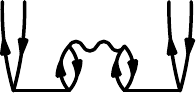}}
  & +
  & \diagramBox[1]{\includegraphics{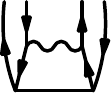}}
  & +
  & \ldots
  \\[6ex]
    t^{ab}_{ij}
  & =
  & (
  & v^{ab}_{ij}
  & +
  & v^{ak}_{ic} t^{cb}_{kj}
  & +
  & v^{kb}_{cj} t^{ac}_{ik}
  & +
  & v^{ab}_{cd} t^{cd}_{ij}
  & +
  & v^{kl}_{cd} t^{ac}_{ik} t^{db}_{lj}
  & -
  & v^{kb}_{id} t^{ad}_{kj}
  & +
  & \ldots
  & )
  & /
  & \Delta^{ij}_{ab}
\end{array}
\label{eq:amplitude}
\end{equation}
\end{widetext}
where $\Delta^{ij}_{ab}=\varepsilon_i+\varepsilon_j-\varepsilon_a-\varepsilon_b$
are the one-electron energy differences in the Hartree--Fock approximation.
The amplitude equation above illustrates that
coupled cluster methods perform a resummation of certain diagrams
including particle-hole ring and particle-particle ladder (ppl) diagrams
to infinite order coupling different diagrammatic channels.
We note that single particle-hole excitation amplitudes ($t_i^a$) 
make no contribution to the wavefunction and correlation energy
of the uniform electron gas~\cite{Shepherd2014}.
Given the amplitudes $t_{ij}^{ab}$ that solve the above equation we can
replace $t^{ab}_{ij}$ in \Eq{eq:gofr} by the right-hand-side
of \Eq{eq:amplitude} to arrive at a
decomposition of the transition
pair correlation function into corresponding diagrammatic contributions
\begin{multline}
g^{\mathrm{CCSD}}(\vec r_{12}) =
  \\ g^\mathrm{driver}(\vec r_{12}) + g^\mathrm{ring}(\vec r_{12}) 
       + g^\mathrm{ppl}(\vec r_{12}) + g^\mathrm{rest}(\vec r_{12}).
\label{eq:gofrdecomp}
\end{multline}
Analogously, the transition structure factor $S^\mathrm{CCSD}(\vec q)$
can be decomposed into its diagrammatic contributions.
We note that a similar labeling of terms in the amplitude equations was
employed in Refs.~\cite{Shepherd2014,Shepherd2014a,bishop_1978,bishop_1982,bishop_1991}.
In contrast to previous work, we perform a decomposition only
in one order of perturbation of the otherwise fully coupled amplitudes.

\begin{figure*}[t]
\begin{center}
\includegraphics{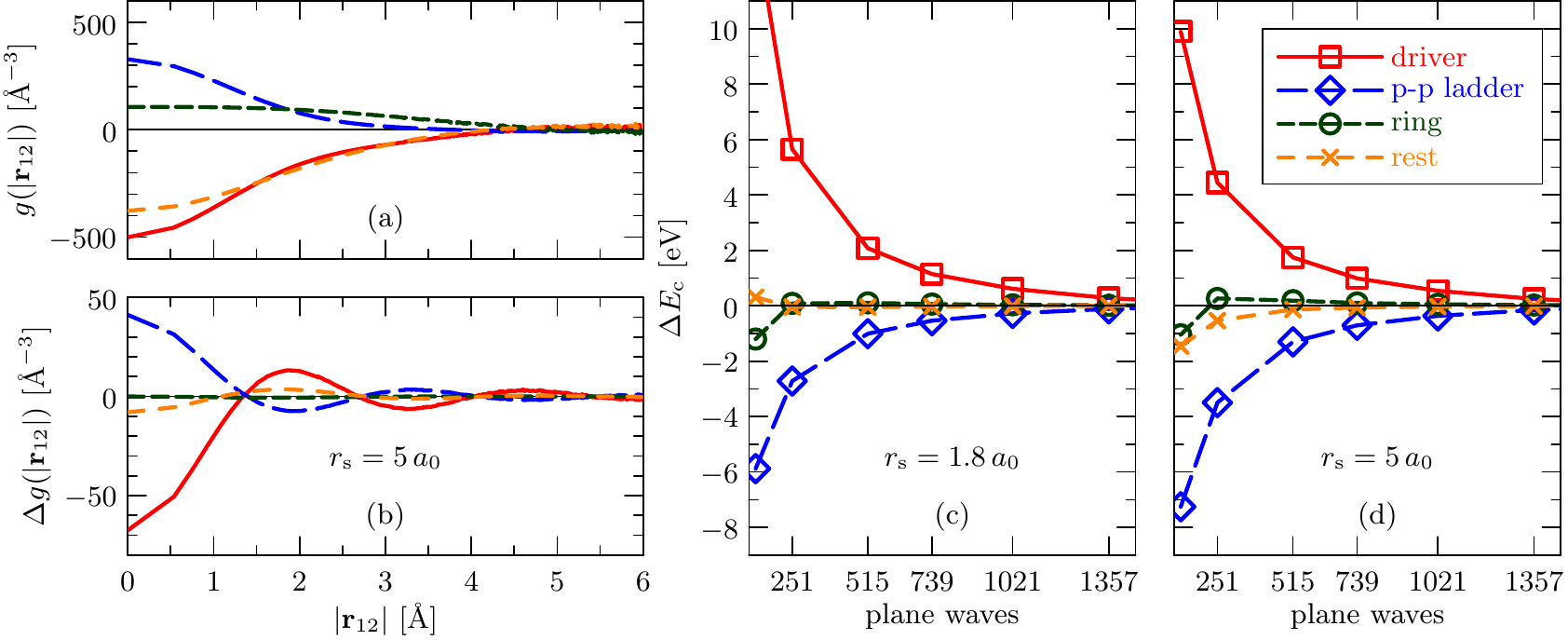}
\end{center}
\caption{%
  (a) Pair correlation function contributions according to \Eq{eq:gofrdecomp}
      using 1863 plane waves at $r_\mathrm{s}=5\,a_0$.
  (b) Difference between pair correlation function contributions obtained using 1863 plane waves and
      1021 plane waves.
  (c)--(d) CBS limit error of corresponding correlation energy contributions
           retrieved as a function of the plane wave basis set size for densities
           corresponding to $r_\mathrm{s}=1.8\,a_0$ and $r_\mathrm{s}=5\,a_0$, respectively.
}
\label{fig:UegChannels}
\end{figure*}

\emph{Results.} -- In the following we will discuss numerical results
of the decomposed pair correlation function
obtained for the UEG. 
The results shown in \Fig{fig:UegChannels}(a) have been computed
employing a 54 electron gas simulation cell at a density corresponding
to $r_\mathrm{s}= 5\,a_0$ identical to the system used in
Ref.~\cite{Grueneis2013}. 
The amplitudes are expanded in an orbtial basis composed of
1863 plane waves.
We first discuss $g^\mathrm{driver}(\vec r_{12})$, which
captures the
second-order correlation energy contribution to \Eq{eq:ecorr}
and exhibits a minimum at $\vec r_{12}=\vec 0$.
Due to the employed finite one-electron basis set approximation,
$g^\mathrm{driver}(\vec r_{12})|_{\vec r_{12} = \vec 0}$ lacks
a derivative discontinuity as required by the cusp condition
\cite{kimball1973}.
In passing we note that
$g^\mathrm{driver}(\vec r_{12}) $
of a single electron pair 
can be well approximated in the complete basis set (CBS) limit
at short interelectronic distances
using a Slater-type correlation function,
$-\gamma^{-1}\e^{-\gamma |\vec r_{12}|}$, 
where $\gamma$ is a parameter that increases with
increasing density~\cite{Grueneis2013}.
In decreasing order of absolute magnitude,
the largest contributions to $g^{\mathrm{CCSD}}(\vec 0)$
originate from $g^\mathrm{driver}(\vec r_{12}) $, $g^\mathrm{rest}(\vec r_{12})$,
$g^\mathrm{ppl}(\vec r_{12})$ and $g^\mathrm{ring}(\vec r_{12})$.
In contrast to $\vec r_{12} = \vec 0$, we find that
the relative contribution of
$g^\mathrm{ring}(\vec r_{12} )$ to $g^{\mathrm{CCSD}}(\vec r_{12} )$ increases with
increasing interelectronic distance,
whereas $g^\mathrm{ppl}(\vec r_{12})$ decays to zero rapidly.
This observation demonstrates that the employed decomposition
allows for an analysis of the interelectronic correlation strength
and its dependence on the distance for individual diagrammatic
contributions.
We also note that the importance of ring diagrams for the
correlation energy at large interelectronic distances
is known and of particular significance to the correct description
of dispersion interactions in semiconductors using the RPA~\cite{Harl2009,Dobson2006}.

We now discuss the convergence of the decomposed pair correlation
function with respect to the employed orbital basis set.
$g^\mathrm{driver}(\vec r_{12} )$ converges slowly to the complete basis set limit.
This can be seen from $\Delta g^\mathrm{driver}(\vec r_{12} )$
as depicted in \Fig{fig:UegChannels}(b), which
corresponds to the difference of $g^\mathrm{driver}(\vec r_{12} )$ 
calculated using 1021 and 1863 plane wave orbitals.
This analysis demonstrates that the
ppl term converges at a similar rate albeit
with an opposite sign as indicated by 
$\Delta g^\mathrm{driver}(\vec r_{12} )$ and $\Delta g^\mathrm{ppl}(\vec r_{12})$ in
\Fig{fig:UegChannels}(b).
Moreover, 
$\Delta g^\mathrm{driver}(\vec r_{12} )$ and $\Delta g^\mathrm{ppl}(\vec r_{12})$ are
largest in magnitude for short interlectronic distances,
illustrating that the increasing basis set is required to capture
short-ranged electronic correlation effects.
Note that other than $g^\mathrm{driver}(\vec r_{12})$
and $g^\mathrm{ppl}(\vec r_{12} )$, 
the ring diagrams converge rapidly
with respect to the employed number of orbitals.
Likewise we find that $g^\mathrm{rest}(\vec r_{12} )$ converges
significantly faster than
$g^\mathrm{driver}(\vec r_{12} )$ and $g^\mathrm{ppl}(\vec r_{12} )$
and we will not perform
a more detailed analysis of this term in the present work.

Based on the decomposition of the electron pair correlation function introduced
above, we now discuss the corresponding correlation energy contributions and
their convergence to the complete basis set limit.
To this end we have estimated the CBS limit reference energies by employing
1863 virtual orbitals, which provides sufficiently well converged values for the
purpose of the present discussion.
We study an identical simulation cell to the one described above
with a density corresponding to $r_\mathrm{s}= 5\,a_0$. In addition
we discuss results obtained for $r_\mathrm{s}= 1.8\,a_0$.
Figures~\ref{fig:UegChannels}(c) and (d) show the convergence of
the basis set incompleteness errors for the decomposed correlation energy 
retrieved as a function of the employed
orbital basis set size
($\Delta E^\mathrm{driver}$, $\Delta E^\mathrm{ring}$, $\Delta E^\mathrm{ppl}$
and $\Delta E^\mathrm{rest}$).
A comparison between \Fig{fig:UegChannels}(c) and
\Fig{fig:UegChannels}(d) reveals that the convergence
behavior of most terms is similar for both densities.
Only $\Delta E^\mathrm{rest}$ decays to zero in one case from above and
in the other case from below.
We find that the slowest decay of the  basis set incompleteness error
is observed for $\Delta E^\mathrm{driver}$ and $\Delta E^\mathrm{ppl}$, 
in agreement with the discussion of the convergence of
the corresponding pair correlation functions carried out in the paragraph above.
$\Delta E^\mathrm{ring}$ and $\Delta E^\mathrm{rest}$ contributions
exhibit a rapid convergence. 

To better understand the reason for the
convergence behavior of
the ring and particle-particle ladder correlation energy contributions
with respect to the employed basis, we perform an
analysis of a simplified system.
To this end we consider two electrons in a box with a homogeneous
background charge such that $\vec k_i=\vec k_j=\vec 0$,
$\vec k_a=-\vec k_b=\vec q$ due to momentum
conservation and the Coulomb integrals are
$v^{ab}_{ij}=4\pi/\vec q^2$.
We separate the kinetic and the exchange part of
one-electron energy differences
$\Delta^{ij}_{ab} = -(\vec q^2+\delta_q^2)$,
where the latter is negligible compared to the kinetic term for
large magnitudes of $\vec q$.
For the amplitudes $t_{ij}^{ab}$
we use a first-order approximation
that reads $t_{ij}^{ab} \approx
-4 \pi/(\vec q^2+\gamma^2)^2$,
which corresponds to the Fourier
transform of the Slater-type correlation function
$-\gamma^{-1} \e^{-\gamma  |\vec r_{12} |}$.
Replacing $t_{ij}^{ab}$ in the expressions for
$S^\mathrm{ring}(\vec q) $ and $S^\mathrm{ppl}(\vec q) $ by the above
approximation in the limit of a large simulation cell yields
\begin{align}
S^{\rm ring}(\vec q) & \approx
\frac{1}{\vec q^2+\delta_q^2}
\left [ \frac{4\pi}{{\vec q}^2}\, \frac{4 \pi}{({\vec q}^2+\gamma^2)^2} \right ]
\label{eq:Sring}{\mathrm{~and}}
\\
S^{\rm ppl}(\vec q ) & \approx
\frac{1}{\vec q^2+\delta_q^2}
\left [ \int\d\vec q'\
\frac{4 \pi}{{(\vec q'-\vec q)}^2}\,
\frac{4 \pi}{({\vec q'}^2+\gamma^2 )^2}
\right ]. \label{eq:Sppl}
\end{align}
Note the duality relation of the terms in brackets:
one is a point-wise product, the other a convolution
of the Coulomb potential with the Slater-type correlation function.
The latter turns into a point-wise product in real space
$\gamma^{-1}\e^{-\gamma |\vec r_{12} |}/|\vec r_{12}|$,
which is the Yukawa potential scaled by $\gamma^{-1}$
whose Fourier transform is $\gamma^{-1}4\pi/(\vec q^2+\gamma^2)$.
The important consequences of the above analysis are as follows:
(i) the asymptotic behavior of the ring transition structure factor is
$\lim_{|\vec q| \rightarrow \infty}S^{\rm ring}(\vec q ) \propto
1/\vec q^8$ , while (ii) the
ppl transition structure factor decays as
$\lim_{|\vec q| \rightarrow \infty}S^{\rm ppl}(\vec q ) \propto
1/\vec q^4$ only, which is identical to
$\lim_{|\vec q| \rightarrow \infty} 
S^{\mathrm{driver}}(\vec q) \propto 1/\vec q^4$.
This explains qualitatively the observed correlation
energy convergence in \Fig{fig:UegChannels}
for the ring and ppl term with respect to
the employed plane wave basis. Moreover, the above
findings indicate that the ppl contribution
becomes more important with decreasing electron density because
$E^{\mathrm{ppl}} \propto \gamma^{-1}$ and $\gamma\propto 1/r_\mathrm{s}$~\cite{grueneis2015}.
The above analysis also shows that the ppl
contribution modifies the first-order coefficient in the Taylor expansion
of $g(|\vec r_{12}|)$ with respect to $|\vec r_{12}|$ around
$|\vec r_{12}| = 0$ from above.

We now propose a finite basis set correction method for coupled cluster theory
that is based on the analysis of the ppl term discussed above.
Coupled cluster methods are becoming increasingly
popular to perform \emph{ab-initio} studies of solids and surfaces
with high accuracy \cite{gruber18b,tsatsoulis2018,tsatsoulis2017,mcclain_2017,kubas2016,boese2016,Yang640,Booth2013},
demonstrating significant potential
to further expand the scope of electronic structure theory calculations in; for example,
the field of surface chemistry or the study of thermodynamic stabilities of solids.
However, the underlying
computational cost is much larger than that of the more
efficient yet less accurate approximate density functional theory
calculations. The main source
of the computational cost in coupled cluster calculations
originates from the ppl term.
Its computational
complexity scales as $\mathcal{O}(N_\mathrm v^4 N_\mathrm o^2)$
in a canonical formulation, where $N_\mathrm v$ and $N_\mathrm o$
are the number of virtual and occupied orbitals, respectively.
However, \Fig{fig:UegChannels} and the discussion above
show that the basis set incompleteness errors of the second-order correlation
and the ppl term
are propotional to each other albeit with an opposite sign.
This observation motivates the following approximation to
the CCSD correlation energy in the CBS limit that aims at accounting
for the basis set incompleteness
error in the  driver ($E^{\mathrm{driver}}$)
and ppl ($E^{\mathrm{ppl}}$) term
explicitly:
\begin{multline}
E^{\mathrm{CCSD-PPL}}_{\mathrm{c}}(N) \\=
E^{\mathrm{CCSD}}_{\mathrm{c}}(N)+
\Delta E^{\mathrm{driver}}(N)+ \Delta E^{\mathrm{ppl}}(N)\,,
\end{multline}
where
\begin{align}
	\Delta E^{\mathrm{driver}}(N) &=
  E^{\mathrm{driver}}(\mathrm{CBS}) - E^{\mathrm{driver}}(N)\,, \\ 
	\Delta E^{\mathrm{ppl\phantom{ve}}}(N) &=
  \frac{\Delta E^{\mathrm{driver}}(N)}{E^{\mathrm{driver}}(N)}\,
  E^{\mathrm{ppl}}(N)\,.
\end{align}
We refer to the employed orbital basis set size by $N$.
$E^{\mathrm{driver}}(\mathrm{CBS})$ can in practice be estimated
in a computational efficient manner using basis set extrapolation or explicit
correlation techniques~\cite{grueneis2015,shepherd2012,schaefer_2017}.

\begin{figure}[t]
\begin{center}
\includegraphics{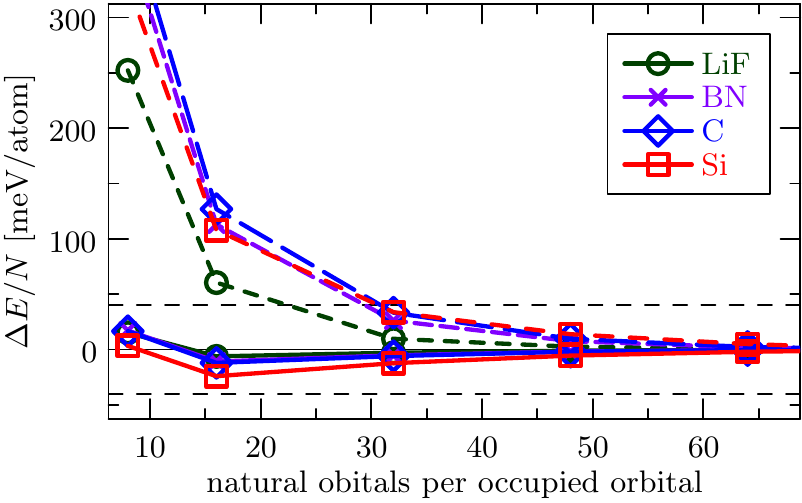}
\end{center}
\caption{%
  CBS limit errors of CCSD (dashed) and CCSD-PPL (solid)
  correlation energies for \emph{ab-initio} systems
  with respect to the number natural orbitals
  employing 8-atomic supercells.
}
\label{fig:AbinitioConvergence}
\end{figure}

\begin{table}
\caption{
CCSD and CCSD-PPL valence electron correlation energies for the Ne atom in m$E_h$.
The CBS limit value is -315.7\,m$E_h$~\cite{Shiozaki2008}.
DZ, TZ and QZ refer to the employed atom-centered Gaussian basis sets aug-cc-pVXZ (X=D,T and Q).
The MP2 CBS limit energy was computed using a [Q5] extrapolation.
}
\label{tab:ecorrval}
\begin{ruledtabular}
\begin{tabular}{l ccc }
                                   &    DZ   &    TZ   &    QZ  \\ \hline
CCSD                               & -210.15 & -274.09 & -297.76  \\
CCSD-PPL                           & -308.08 & -314.95 & -317.27  \\
\end{tabular}
\end{ruledtabular}
\end{table}

We now assess the efficiency of the proposed finite basis set
correction to the coupled cluster correlation energy for
\emph{ab-initio} systems including solids and atoms.
All periodic calculations of solids have been performed using
the Vienna \emph{ab-initio} simulation package (\texttt{VASP})~\cite{kresse1996}
in the framework of the projector augmented wave method~\cite{blochl1994},
interfaced to our coupled cluster code~\cite{hummel_2017} that employs an
automated tensor contraction framework (\texttt{CTF})~\cite{solomonik2014}.
We use natural orbitals to achieve a compact approximation for the virtual
orbital space~\cite{gruneis11}.
\Fig{fig:AbinitioConvergence} depicts the coupled cluster
singles and doubles correlation energy convergence for a range of
periodic crystals: C (diamond), BN (wurtzite), LiF (rock-salt) and Si (diamond),
calculated with and without the proposed finite basis set correction.
Our findings unequivocally demonstrate that the basis set convergence of the coupled
cluster singles and doubles correlation energy is significantly faster including the 
proposed correction compared to uncorrected CCSD energies.
Already 10 natural orbitals per occupied orbital agree with the CBS limit results
to within chemical accuracy ($\approx$40\,meV/atom) for the total energy per atom
as indicated by the region between the dashed horizontal lines.
80 natural orbitals per occupied orbital serve as the approximate CBS reference
in the present work.
In contrast to CCSD-PPL, CCSD requires approximately 30 natural orbitals per occupied
orbital to achieve a similar level of precision.
We stress that a reduction in the number of virtual orbitals by a factor three,
reduces the computational cost for the ppl term by two orders of magnitude.

To further assess CCSD-PPL, we have carried out calculations for the Ne atom
employing a modified version of an open-source quantum chemistry code (\texttt{PSI4}~\cite{PSI4ref})
and atom-centered Gaussian basis sets.
Table~\ref{tab:ecorrval} shows that CCSD-PPL converges rapidly to the CBS limit.
Employing a double-$\zeta$ (DZ) basis
set yields CCSD-PPL correlation energies that are closer to the CBS limit than CCSD correlation energies
obtained using a quadruple-$\zeta$ (QZ) basis set.
Therefore the conclusions drawn from the uniform electron gas also
hold in the atomic limit. We refer the reader to Ref.~\cite{Irmler2019b} for
a more extensive investigation of CCSD-PPL in atoms and molecules
and a comparison to explicitly correlated methods.

\emph{Concluding Remarks.} -- We have presented a diagrammatic decomposition of the
coupled cluster pair correlation function and performed an
analysis of the ring and particle-particle ladder terms.
Based on their dual structure and the observed convergence
with respect to the employed basis, we have introduced an efficient finite
basis set correction that allows for reducing the computational cost
of coupled cluster theory calculation in atoms, molecules and solids substantially.
In combination with the recently proposed finite size corrections \cite{Gruber2018}
and other techniques~\cite{grueneis2015,hummel_2017,booth2016}, this
paves the way for a routine use of coupled cluster theory
in electronic structure calculations of solids and surfaces.
However, the significance of the present work extends far beyond the
proposed basis set correction.
Given  exact reference results for the spin-resolved pair correlation function
in \emph{ab-initio} or model systems that can be obtained using;
for example, full configuration interaction quantum Monte Carlo~\cite{Booth2013},
the errors in the diagrammatic channels
of approximate many-electron theories can be analysed in detail,
allowing for developing more accurate and better balanced truncations
to the many body perturbation expansion of the exact electronic
correlation energy.
%
Finally, we note that the introduced decomposition
scheme can also help developing more efficient embedding and local correlation theories,
where the decoupling and decay of interelectronic correlation effects
in different diagrammatic channels plays a crucial role and determines
the accuracy of the employed approximations.

\emph{Acknowledgements.} --  
The authors thankfully acknowledge support and funding from the European
Research Council (ERC) under the European Unions
Horizon 2020 research and innovation program (Grant Agreement No 715594).
A.~G.~gratefully acknowledges many helpful discussions on F12 theory with
Seiichiro Ten-no and David Tew.

\bibliography{PPL}

\end{document}